\documentclass[11pt,twoside]{article}
\usepackage{asp2010}

\resetcounters

\begin{document}

\title{The mass-loss dominated lives of the most massive stars}
\author{Jorick S. Vink$^1$ 
\affil{$^1$Armagh Observatory, College Hill, BT61 9DG Armagh, UK}}

\begin{abstract}
Utrecht has a long tradition in both spectroscopy and mass-loss studies. 
Here we present a novel methodology to 
calibrate mass-loss rates on purely {\it spectroscopic} 
grounds. We utilize this to predict the final fates of massive stars, involving 
pair-instability and long gamma-ray bursts (GRBs) 
at low metallicity $Z$.
\end{abstract}

\section{Introduction}

Mass loss is an important ingredient in massive star evolution modelling. 
Nowhere is it more dominant than for the most massive stars, which are 
thought to evolve chemically homogeneously (Gr\"afener et al. 2011). 
Mass loss determines the final fate. This may 
involve normal supernovae (SNe) or pair-instability SNe, 
leaving either normal black holes or intermediate mass-black 
holes (IMBHs) behind 
(Belkus et al. 2007; Yungelson et al. 2008). 
The point is that very massive stars (VMS) up to 300$M_{\odot}$ 
-- and perhaps even as high as $\sim$1000$M_{\odot}$ through collisions in 
clusters -- are now thought to exist in nature
(Crowther et al. 2010; Bestenlehner et al. 2011). 

In recent years a debate has arisen regarding the roles of stellar wind 
versus eruptive mass loss, as winds have been found to be clumped, which resulted
in the reduction of empirical mass-loss rates. 
Stellar evolution modellers however generally 
employ theoretical mass-loss rates. These 
are already reduced by a moderate factor (of $\sim$2-3) compared 
to non-corrected empirical rates. 
A key question is whether these reduced rates are correct 
or if they need to be reduced even more. 

Whilst stationary winds in O and Wolf-Rayet (WR) stars are ubiquitous, it is 
not clear if objects like $\eta$\,Car have experienced 
a special evolution, e.g. involving a merger, or if a majority of 
massive stars would encounter eruptive mass-loss phases. 
Alternatively, for the most massive main-sequence WNh stars
there is strong evidence for a Eddington parameter 
$\Gamma$-dependent mass loss (Gr\"afener et al.  2011). 
In other words, for VMS the role of stationary mass loss has 
increased in recent years.

We recently introduced 
the transition mass-loss rate $\dot{M}_{\rm trans}$ between O and WR stars (Vink \& Gr\"afener 2012). 
Its novelty is that it is model independent. 
All that is needed is the {\it spectroscopic} transition point in a data-set, and 
to determine the stellar luminosity. This is far less model-dependent than conventional 
mass-loss diagnostics. 
Our results suggest that the rates provided by Vink et al. (2000) 
are of the right order of magnitude in the ~50$M_{\odot}$ range, but 
alternative mechanisms might be needed at lower $Z$, particularly relevant for the occurrence 
of GRBs.

\section{The mass loss versus Eddington-Gamma dependence}

\begin{figure}
\begin{center}
 \includegraphics[width=3.4in]{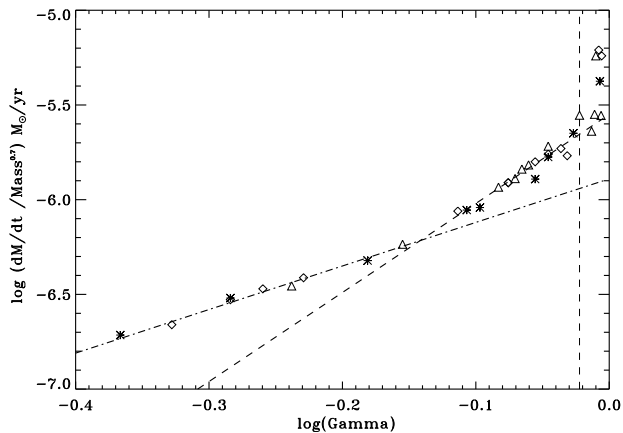} 
\caption{Mass-loss predictions versus the Eddington parameter $\Gamma$ -- divided by $M^{0.7}$. 
Symbols correspond to models of different mass ranges (Vink et al. 2011a).}
\label{f_mdotvink}
\end{center}
\end{figure}

Vink et al. (2011a) discovered a kink in the slope of the
mass-loss vs. $\Gamma$ relation at the transition from optically thin O-type
to optically thick WR-type winds. 
Figure\,\ref{f_mdotvink} depicts mass-loss predictions for VMS 
as a function of the Eddington parameter $\Gamma$. 
For ordinary O stars with ``low'' $\Gamma$ the $\dot{M}$ $\propto$ 
$\Gamma^{x}$ relationship is shallow, with $x$ $\simeq$2. 
There is a steepening at higher $\Gamma$, where
$x$ becomes $\simeq$5. Here the optical depths and wind efficiencies exceed 
unity. This result from Monte Carlo modelling, i.e. 
that the O to WR mass loss transition point coincides with $\eta \simeq 1$, 
can also be found analytically. 

\subsection{The transition mass-loss rate}
\label{sec_anal}

Lamers \& Cassinelli (1999) provided
momentum considerations for dust-driven winds 
that we apply to line-driven winds. 
The integral form of the momentum equation has four terms, but  
as hydrostatic equilibrium is a good approximation 
for the inner wind, and the gas pressure 
gradient is small beyond the sonic point, 
the 2nd and 3rd terms are neglected, leading to:

\begin{equation}
\int_{R_{\star}}^{\infty} 4 \pi r^2 \rho v \frac{dv}{dr} dr + \int_{r_{\rm s}}^{\infty} \frac{G M}{r^2} (1 - \Gamma)  \rho 4 \pi r^2 dr = 0.
\label{eq_LC99}
\end{equation}
When we utilize the mass-continuity equation $\dot{M} = 4 \pi r^2 \rho v$, we 
retrieve
\begin{equation}
\int_{R_{\star}}^{\infty} \dot{M} \frac{dv}{dr} dr = \dot{M} v_{\infty} = 4 \pi G M \int_{r_{\rm s}}^{\infty} (\Gamma(r) - 1)\rho dr 
\end{equation}
where $r_s$ represents the sonic radius and $\Gamma(r) = \frac{\kappa_{\rm F} L}{4\pi c GM}$ 
is the Eddington factor with respect to the flux-mean opacity $\kappa_{\rm F}$.
Employing the wind optical depth $ \tau = \int_{r_s}^\infty 
\kappa_{\rm F}\rho\, {\rm d}r$, we obtain

\begin{equation}
\dot{M} v_{\infty} \simeq \frac{4 \pi G M}{\kappa} (\Gamma - 1) \tau = \frac{L}{c} \frac{\Gamma-1}{\Gamma} \tau.
\label{eq_gaga}
\end{equation}
If we assume that $\Gamma$ is larger than unity, the factor
$(\Gamma-1)/\Gamma$ is close to one, which  
leads to $\dot{M} v_{\infty} = L/c \tau$.
We can now derive the unique wind efficiency condition 
$\eta = \frac{\dot{M} v_{\infty}}{L/c} = \tau = 1$ for the transition point
from optically thin O-star winds to optically-thick WR winds, and we 
retrieve a {\it model-independent} $\dot{M}$. 
In case we have an empirical data-set available 
that includes luminosity determinations we can provide 
the transition mass-loss rate $\dot{M}_{\rm trans}$, simply from the transition 
luminosity $L_{\rm trans}$ (and the terminal velocity $v_{\infty}$): 
$\dot{M}_{\rm trans} = \frac{L_{\rm trans}}{v_{\infty} c}$. 
This transition point is obtainable just by spectroscopic means, 
and {\em independent} of any wind-clumping assumptions.

\subsection{Do very massive stars make IMBHs or pair instability SNe?}

To address these questions, we require 
accurate mass-loss rates for VMS.  
As Vink \& Gr\"afener (2012) were able to calibrate the 
Vink et al. (2000) mass-loss prescriptions at the high-mass end, and 
given that these rates agree well with the 
Crowther et al. (2010) rates for the 30\,Dor WNh core stars for moderate 
clumping factors $D\sim10$, we employ the Vink et al. (2000) rates for now. 
When starting with 300$M_{\odot}$, we find $\dot{M}$ $= 10^{-4.2}$ 
$M_{\odot} {\rm yr}^{-1}$.
For a 2.5Myrs lifetime this leads to a total 
main-sequence mass lost of $\simeq$150$M_{\odot}$. 
Extra wind mass loss during the core helium phase will 
further evaporate these stars. There seems   
little space left for eruptive mass loss.
Our results would imply that IMBHs and 
pair-instability SNe are unlikely, unless we move 
to lower $Z$.

\section{GRBs and the metallicity dependence of Wolf-Rayet stars}

\begin{figure}
\begin{centering}
\includegraphics[width=8cm]{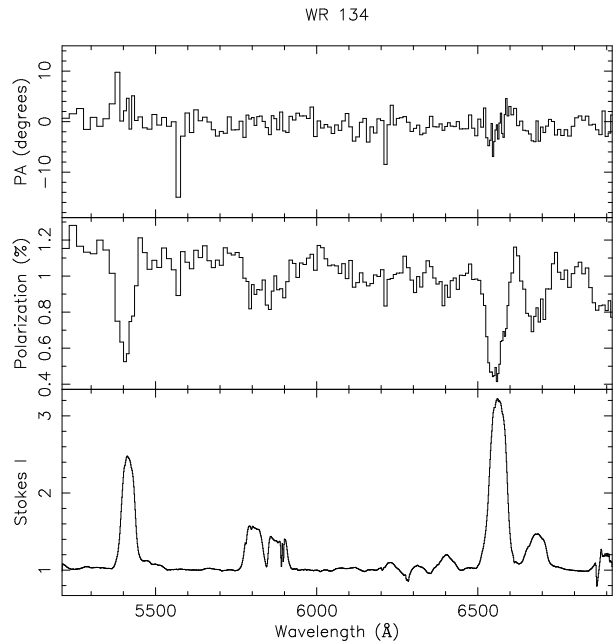}
\caption{Galactic Wolf-Rayet star WR 134. (\textit{a})
Position Angle (PA) of the polarisation. 
 (\textit{b})
Degree of linear polarisation. (\textit{c})  
Stokes I ``intensity''. 
\label{fig2}}
\end{centering}
\end{figure}

Mass loss at low $Z$ has gained attention due to the issue of cosmic reionization by 
population {\sc iii} stars. Also, massive stars are thought to be 
the progenitors of GRBs.
Within the collapsar model, GRB progenitors require 2 properties: (i) 
rapidly rotating cores, and (ii) the absence of hydrogen envelopes. 
Therefore, GRB progenitors are thought to be rotating WR stars. 
A potential pitfall is that WR star have high mass loss which 
could remove all the angular momentum before core collapse. 
In the rapidly rotating stellar models of Yoon \& Langer (2005), 
the objects evolve ``quasi-homogeneously''. 
The objects are subjected to a strong magnetic coupling between 
the stellar core and envelope. If rapid rotation can be maintained 
as a result of lower main-sequence mass loss in lower $Z$ galaxies, the objects may 
avoid spin-down during a RSG or LBV phase, and directly 
become rotating WR stars. 
If WR winds also depend on Fe driving (Vink \& de Koter 2005), the 
objects could remain rapid rotators towards the end of their lives, enabling 
GRB formation, but {\it exclusively} at low $Z$. 

Recent GRB observations however suggest that GRBs are not restricted 
to low $Z$. Indeed, there seems to be a need for a high $Z$ GRB channel. 
We have recently identified a sub-group of rotating Galactic WR 
stars. This would allow for a potential solution to the problem 
(Vink et al. 2011b; Gr\"afener et al. 2012). 
Spectropolarimetry surveys show that the majority of WR stars have spherically 
symmetric winds indicative of slow rotation, but a small minority display 
signatures of a spinning stellar surface (see Fig.\,2 for the example of WR\,134). 
We recently found the spinning subgroup surrounded by ejecta 
nebulae -- thought to be ejected during a recent RSG/LBV phase --  
suggesting that these objects are still young, and rotating. 

If core-surface coupling were strong enough the cores would not 
rotate rapidly enough to make GRBs. However, if core-envelope 
coupling is less efficient stars may have sufficient 
angular momentum in the core to make a GRB. 
At high $Z$, the objects would in most cases 
still spin down due to mass loss, but with our post-RSG/LBV 
scenario one would not exclude the option of high $Z$ GRBs. 
Still, low $Z$ remains preferred due to weaker WR winds
(Vink 2007). 

\acknowledgements I would like to thank all SIU staff 
for their support during the 9 years of my under/graduate 
studies from 1991 to 2000. I would like to thank
Bram, Frank, Henny, Jan, Max and Rob for very stimulating lectures 
in astrophysics.


\end{document}